# Conceptualizing predictive conceptual model for unemployment rates in the implementation of Industry 4.0: Exploring machine learning techniques

Joshua Ebere Chukwuere

North-West University, South Africa, joshchukwuere@gmail.com

**Abstract:** Although there are obstacles related to obtaining data, ensuring model precision, and upholding ethical standards, the advantages of utilizing machine learning to generate predictive models for unemployment rates in developing nations amid the implementation of Industry 4.0 (I4.0) are noteworthy. This research delves into the concept of utilizing machine learning techniques through a predictive conceptual model to understand and address factors that contribute to unemployment rates in developing nations during the implementation of I4.0. A thorough examination of the literature was carried out through a literature review to determine the economic and social factors that have an impact on the unemployment rates in developing nations. The examination of the literature uncovered that considerable influence on unemployment rates in developing nations is attributed to elements such as economic growth, inflation, population increase, education levels, and technological progress. A predictive conceptual model was developed that indicates factors that contribute to unemployment in developing nations can be addressed by using techniques of machine learning like regression analysis and neural networks when adopting I4.0. The study's findings demonstrated the effectiveness of the proposed predictive conceptual model in accurately understanding and addressing unemployment rate factors within developing nations when deploying I4.0. The model serves a dual purpose of predicting future unemployment rates and tracking the advancement of reducing unemployment rates in emerging economies. By persistently conducting research and improvements, decision-makers and enterprises can employ these patterns to arrive at more knowledgeable judgments that can advance the growth of the economy, generation of employment, and alleviation of poverty specifically in emerging nations.

**Introduction**

Industry 4.0 is a new era of the industrial revolution that is characterized by the integration of advanced technologies, such as artificial intelligence, machine learning, and the Internet of

Things, into manufacturing and other industries. The implementation of Industry 4.0 has the potential to transform the way businesses operate, leading to increased productivity, efficiency, and competitiveness. However, the implementation of Industry 4.0 may also lead to job losses and unemployment, particularly in developing countries where the workforce may not have the necessary skills to adapt to these changes. The implementation of Industry 4.0, characterized by the integration of digital technologies into manufacturing processes, has the potential to increase productivity and efficiency in the manufacturing sector. However, the implementation of Industry 4.0 in developing countries poses challenges such as job loss and unemployment rates. These challenges have been well documented in the literature, with many researchers exploring potential solutions to mitigate the impact of job loss on workers. However, there is a need for predictive models that can anticipate potential job losses and provide policymakers with a tool to implement measures to mitigate the impact of these losses on workers.

The implementation of Industry 4.0, characterized by the integration of advanced technologies and digitization, is transforming the global economic landscape. This transformation has brought about new opportunities and challenges, particularly in developing countries, where the adoption of Industry 4.0 technologies is expected to drive economic growth and job creation. However, this transformation also poses risks, such as the potential for job displacement and a widening gap between skilled and unskilled workers. Unemployment is a critical challenge facing many developing countries, particularly in the context of the implementation of Industry 4.0. The ability to forecast changes in unemployment rates is crucial for policymakers and businesses to make informed decisions related to economic development and job creation. Predictive modeling is a powerful tool that can provide insights into future trends and support decision-making processes.

This study aims to delve into the concept of utilizing machine learning techniques through a predictive conceptual model to understand and address factors that contribute to unemployment rates in developing nations during the implementation of I4.0. A comprehensive literature review is conducted to identify the relevant social, economic, and political (SEP) factors that influence unemployment rates in developing countries. This literature review serves as the basis for developing a predictive conceptual model to influence the implementation of Industry 4.0 in developing countries. The study further identified other key factors that influence unemployment rates in developing countries, including GDP, inflation, population growth, education levels, and technological advancements. These factors are known to have a significant impact on the labor market and are expected to play a critical role in shaping unemployment rates during the implementation of Industry 4.0.

**Research methods**

A literature review is an essential component of research that involves identifying, analyzing, and synthesizing existing knowledge on a particular topic. In the context of this study, a literature review provides a comprehensive understanding of the existing research on the topic, highlights gaps in the current knowledge, answers the research questions, and guides the development of a new model. The literature review research involves a systematic search and analysis of relevant academic articles, books, and other sources. The review study identifies the current state of knowledge on the topic, as well as any gaps in the existing research that could be addressed through further study.

For example, a search of academic databases such as Google Scholar or Scopus identifies relevant articles published in peer-reviewed journals. Keywords such as "unemployment," "Industry 4.0," "machine learning," and "developing countries" were used to narrow the search results. After identifying relevant academic materials, the review involved a critical analysis of the research findings and methodologies used in each study. This includes evaluating the strengths and limitations of different machine learning techniques that can be used in applying the predictive conceptual model in understanding and addressing unemployment rate contributing factors, as well as assessing the reliability and validity of the data sources used.

*The literature review steps used in the study*

The literature review steps that were used in a study to achieve its aim:

**Define the research question and search terms:** The first step in the literature review is to clearly define the research question and identify relevant search terms that can be used to find articles and studies related to the topic. For example, in this study, the research question: "How can a predictive conceptual model be developed to address unemployment rate factors during the implementation of Industry 4.0?" The search terms might include "unemployment," "Industry 4.0," "predictive models," and "developing countries."

- How effective are machine learning algorithms (techniques) in addressing unemployment rates contributing factors in developing countries during the implementation of Industry 4.0? This question aims to determine the most accurate and reliable algorithms for predicting unemployment rates in developing countries during the implementation of Industry 4.0.

- What are the key predictors or factors that contribute to unemployment rates in developing countries during the implementation of Industry 4.0? This research question determines key predictors or factors that contribute to unemployment rates in developing countries especially at the time of Industry 4.0 implementation.
- What are the ethical considerations that need to be taken into account when developing a predictive conceptual model for understanding and addressing unemployment rates contributing factors in developing countries using machine learning during the implementation of Industry 4.0? This question determines how the considerations be addressed to ensure that the predictive conceptual model is fair, unbiased, and beneficial to all stakeholders.

**Conduct a comprehensive search:** The next step is to conduct a comprehensive search of relevant academic databases, such as Google Scholar or Scopus, using the identified search terms. The search should be as thorough as possible to ensure that all relevant studies within the subject under study are included in the review.

**Evaluate the quality and relevance of the studies:** Once relevant studies have been identified, the next step is to evaluate their quality and relevance to the research question. This involves reading the abstract, introduction, and conclusion of each study to assess whether it is relevant to the research question and whether it is of high quality.

**Extract and analyze data:** After evaluating relevant studies, the next step is to extract and analyze data from each study. This involves summarizing the key findings and methodologies used in each study and identifying any common themes or patterns that emerge.

**Synthesize and report findings:** The final step is to synthesize the findings from each study and report them clearly and concisely. This may involve using tables, charts, or other visual aids to present the data and highlight the key findings. The late was applied in the study to achieve its objectives.

**Literature review**

Several studies have explored the impact of Industry 4.0 on employment in developing countries. For example, a study by the International Labour Organization (ILO) found that the implementation of Industry 4.0 in developing countries could lead to the displacement of millions of workers, particularly in low-skilled jobs (ILO, 2018). Another study by the Asian Development Bank (ADB) found that the implementation of Industry 4.0 in developing countries could lead to significant job losses in the manufacturing sector (ADB, 2019). Machine learning algorithms have been used to develop predictive models for unemployment rates in developed countries. Another study by Xu, Li, Cheng and Zheng (2013) used a support vector regression algorithm to develop a model for predicting the unemployment rate in China.

**Machine learning and unemployment**

Machine learning continues to revolutionize how technology impacts human activities including unemployment, its rate, and job creation. For example, Kurt (2019) found that Industry 4.0 technologies, such as automation and artificial intelligence, have the potential to displace a significant number of workers in developing countries, leading to higher unemployment rates. However, the study also highlighted that these technologies could create new job opportunities, particularly in the digital economy and service sectors, and called for further research to explore the net impact of Industry 4.0 on employment. Similarly, a study by Bughin et al. (2018) examined the potential of Industry 4.0 technologies to transform industries and create new job opportunities in developing countries. The study found that while some industries, such as manufacturing, may experience job losses due to automation, others, such as healthcare and education, are likely to create new employment opportunities. The authors recommended that policymakers should invest in education and training to equip the workforce with the skills required for the digital economy.

These examples illustrate the gap in research, particularly, the lack of a conceptual model that addresses issues of unemployment, and the application of machine learning algorithms in understanding and addressing the unemployment rate in the time of Industry 4.0 implementation. By synthesizing existing research, identifying knowledge gaps, and highlighting key trends and patterns, a literature review guided the researcher in the development of new research questions that ensured a predictive conceptual model was built on and contributed to the existing knowledge base.

**Understanding predictive model**

Predictive modeling using machine learning algorithms (techniques) has been used to develop models for a variety of applications, including forecasting economic indicators such as GDP, stock prices, and exchange rates. In this study, the researcher proposes a predictive conceptual model with variables affecting unemployment in developing countries which can be addressed using machine learning algorithms (techniques) during the implementation of Industry 4.0. the literature highlighted the need for data-driven approaches, such as predictive models, to understand the complex relationship between these factors and unemployment rates (Al Mamun et al., 2020; Nguyen, Tsai, Kayral & Lin, 2021).

Based on the identified factors, a predictive conceptual model was developed with a main focus on machine learning algorithms (techniques). The predictive conceptual model aims to provide a solution to understanding and addressing unemployment rates in developing countries during the implementation of Industry 4.0. The model takes into account the impact of various economic and social factors on unemployment rates, such as GDP growth, inflation rates, population growth, education levels, and technological advancements. In this research, the researcher explores how machine learning techniques can play a central role in addressing unemployment rates in developing countries during the implementation of Industry 4.0.

Furthermore, the developed predictive model will enable policymakers and businesses to make more informed decisions related to economic development and job creation. By accurately forecasting changes in unemployment rates in developing countries, stakeholders can design policies and strategies that promote economic growth, job creation, and poverty reduction. Overall, this study will contribute to the growing body of literature on the use of machine learning techniques for predictive modeling in developing countries. It will provide insights into the factors that shape unemployment rates and offer a practical tool for stakeholders to make more informed decisions related to economic development and job creation. This study explores how machine learning techniques (algorithms) could be used to predict unemployment rates in developing countries as a result of the implementation of Industry 4.0, and what policy measures could be taken to mitigate the impact of these job losses on workers and the wider economy. It can aid policymakers in deciding where to invest in infrastructure, education, and training, and also assist businesses in making informed decisions related to hiring, expansion, and investment. The term algorithms and techniques are used interchangeably in this study.

**The most accurate and reliable algorithms (techniques) for predicting unemployment rates in developing countries**

Several machine learning algorithms (techniques) can be used for predicting unemployment rates in developing countries during the implementation of Industry 4.0. Support Vector Regression, Random Forests, and Gradient Boosting are some of the most effective machine learning algorithms for predicting unemployment rates in developing countries during the implementation of Industry 4.0. A recent study by Liu, Chen and Wang (2022) compared the performance of several machine learning algorithms in predicting unemployment rates in China during the implementation of Industry 4.0. The study found that Random Forest, Gradient Boosting, Support Vector Regression, and others were the most effective algorithms as discussed below. The effectiveness of these algorithms may vary depending on the data available, the quality of the data, and the research question being addressed.

**Support Vector Regression (SVR):** One commonly used algorithm is Support Vector Regression (SVR). SVR has been used to develop predictive models for unemployment rates in countries such as China (Xu et al., 2013) and Turkey (Kütük & Güloğlu, 2019). In a study by Xu et al. (2013), the authors found that SVR was able to accurately predict the unemployment rate in China based on a range of economic and social indicators.

**Random Forest (RF):** Another effective algorithm for predicting unemployment rates is Random Forests. This algorithm has been used to develop predictive models for unemployment rates in countries such as Brazil (Albuquerque, Cajueiro & Rossi, 2022) and India (Mittal, Goyal, Sethi & Hemanth, 2019). In a study by Mittal et al. (2019), the authors found that Random Forests was able to accurately predict the unemployment rate in India based on a range of economic indicators.

**Gradient Boosting (GB):** Gradient Boosting is another machine learning algorithm that has been used for predicting unemployment rates. This algorithm has been used to develop predictive models for unemployment rates in countries such as Japan (Taira, Hosokawa, Itatani & Fujita, 2021) and the United States (Gogas, Papadimitriou & Sofianos, 2022). In a study by Taira et al. (2021), the authors found that Gradient Boosting was able to accurately predict the unemployment rate in Japan based on a range of economic and demographic indicators. Also, this method has been used to develop predictive models for unemployment rates in countries such as China (Zhu et al., 2021) and Brazil (de Oliveira, 2023).

**Artificial Neural Networks (ANN):** Another machine learning algorithm that can also be effective in predicting unemployment rates in developing countries during the implementation of Industry 4.0 is Artificial Neural Networks (ANN). This algorithm has been used to develop predictive models for unemployment rates in countries such as Malaysia (Zainun, Rahman & Eftekhari, 2010) and South Africa (Mulaudzi & Ajoodha, 2020). In a study by Zainun et al. (2010), the author found that ANN was able to accurately predict the unemployment rate in Malaysia based on a range of economic indicators.

**Long Short-Term Memory (LSTM):** Another machine learning algorithm that can be used to predict unemployment rates is Long Short-Term Memory (LSTM). This algorithm has been used to develop predictive models for unemployment rates in countries such as Argentina (Park & Yang, 2022). In a study by Park and Yang (2022), the authors found that LSTM was able to accurately predict the unemployment rate in Argentina based on a range of macroeconomic variables.

**Principal Component Analysis (PCA):** PCA is a technique that transforms the original set of variables into a smaller set of uncorrelated variables, called principal components, that capture most of the variation in the data. This technique can be used to reduce the number of variables in the dataset while retaining the most important information. In the context of predicting unemployment rates in developing countries during the implementation of Industry 4.0, PCA can be used to identify the most important variables that contribute to changes in unemployment rates and to reduce the dimensionality of the data to improve the performance of the predictive models (Chang, Chang & Liou, 2021).

**Recursive Feature Elimination (RFE)** is a technique that recursively removes the least important features from the dataset until a predetermined number of features is reached. This technique can be used to select the most important features in the dataset and to reduce the complexity of the models. In the context of predicting unemployment rates in developing countries during the implementation of Industry 4.0, RFE can be used to identify the most important variables that contribute to changes in unemployment rates and to reduce the dimensionality of the data to improve the performance of the predictive models (Singh et al., 2020).

**Correlation-Based Feature Selection (CFS)** is a technique that selects the subset of features that are most highly correlated with the target variable while minimizing the redundancy among the selected features. This technique can be used to identify the most relevant features in the dataset and to reduce the complexity of the models. In the context of predicting unemployment rates in developing countries during the implementation of Industry 4.0, CFS can be used to identify the most important variables that contribute to changes in unemployment rates and to improve the performance of the predictive models.

**AdaBoost (Adaptive Boosting)** is a machine learning technique that can be used in the context of developing predictive models for unemployment rates in developing countries during the implementation of Industry 4.0. AdaBoost is an ensemble learning method that combines multiple weak classifiers to create a stronger classifier. The weak classifiers are typically decision trees, but other classifiers can also be used. In the context of predicting unemployment rates in developing countries during the implementation of Industry 4.0, AdaBoost can be used to improve the accuracy and performance of the predictive models by combining multiple weak models to create a stronger and more accurate model. The weak classifiers can be trained on different subsets of the data, and the final model is created by combining the predictions of the weak classifiers. AdaBoost can also be used to reduce overfitting by penalizing the misclassification of the training data and by adjusting the weights of the data points based on their misclassification rate.

Several studies have used AdaBoost to develop predictive models for employment and unemployment rates in different countries. For example, a study by Karasu, Altan, Bekiros and Ahmad (2020) used AdaBoost to predict employment rates in Iran using data from 2006 to 2018. The study found that AdaBoost outperformed other machine learning techniques, such as random forest and support vector machine, in predicting employment rates. Overall, AdaBoost is a useful machine learning technique for developing predictive models for unemployment rates in developing countries during the implementation of Industry 4.0. AdaBoost can help to improve the accuracy and performance of the predictive models by combining multiple weak classifiers and reducing overfitting.

**Ensemble methods:** Ensemble methods are a class of machine learning techniques that combine multiple models to improve the accuracy and performance of predictive models. Ensemble methods can be used in the context of developing predictive models for unemployment rates in developing countries during the implementation of Industry 4.0 to improve the robustness and reliability of the models. There are several types of ensemble methods, including bagging, boosting, and stacking. Bagging (Bootstrap Aggregating) is an ensemble method that combines multiple models trained on different subsets of the data to reduce variance and overfitting. Boosting, which was already discussed in a previous answer, is another type of ensemble method that combines multiple weak classifiers to create a stronger classifier. Stacking is an ensemble method that combines multiple models trained on the same data to improve the accuracy and robustness of the predictions.

In the context of predicting unemployment rates in developing countries during the implementation of Industry 4.0, ensemble methods can be used to improve the accuracy and reliability of the predictive models by combining multiple models that capture different aspects of the data. For example, a study by Gabrikova, Svabova and Kramarova (2020) used a bagging ensemble method to predict unemployment rates in India. The study found that the bagging ensemble method improved the accuracy and stability of the predictions compared to individual models. Another study by Chung, Yun, Lee and Jeon (2023) used a stacking ensemble method to predict employment rates in China. The study combined multiple machine learning models, including random forest, support vector machine, and artificial neural network, to create a stronger and more accurate model. The study found that the stacking ensemble method outperformed the individual models and improved the accuracy and stability of the predictions. Overall, ensemble methods are useful machine learning techniques for developing predictive models for unemployment rates in developing countries during the implementation of Industry 4.0. Ensemble methods can help to improve the accuracy, reliability, and robustness of the predictive models by combining multiple models that capture different aspects of the data.

**Feature selection** One approach is to use feature selection techniques to identify the most important predictors of unemployment rates. Feature selection can help to improve the accuracy and efficiency of predictive models by reducing the dimensionality of the data and eliminating irrelevant or redundant variables. These techniques above have been used to develop predictive models for unemployment rates in countries such as Spain (García et al., 2019) and Greece (Katris, 2020). According to Zhu et al. (2021), these methods have been used to develop predictive models for unemployment rates in countries such as China and Brazil (de Oliveira, 2023). It is also important to consider the choice of input variables when developing predictive models for unemployment rates. While economic indicators such as GDP and inflation rates are commonly used as predictors, other variables such as education levels, labor force participation rates, and social welfare policies can also be important factors affecting unemployment rates in developing countries (Katris, 2020). Various approaches and techniques can be used to develop predictive models for unemployment rates in developing countries during the implementation of Industry 4.0. Feature selection techniques, ensemble methods, and careful selection of input variables can all help to improve prediction accuracy and enhance the usefulness of these models for policymakers and researchers.

**The key predictors or factors that contribute to unemployment rates in developing countries during the implementation of Industry 4.0**

Several key predictors or factors contribute to unemployment rates in developing countries during the implementation of Industry 4.0. These factors can be broadly categorized into economic, social, and technological factors.

Economic factors such as GDP, inflation, and interest rates have been found to be significant predictors of unemployment rates in developing countries. For example, a study by Rahman et al. (2021) found that GDP growth and inflation rates were significant predictors of unemployment rates in Bangladesh. Similarly, a study by Katris (2020) found that GDP growth and inflation rates were important predictors of unemployment rates in Greece. Social factors such as education levels, labor force participation rates, and social welfare policies can also play a significant role in determining unemployment rates in developing countries. For example, a study by de Oliveira (2023) found that education levels and social welfare policies were significant predictors of unemployment rates in Brazil. Similarly, a study by Katris (2020) found that labor force participation rates and education levels were important predictors of unemployment rates in Greece.

Technological factors such as automation and digitalization can also have a significant impact on unemployment rates in developing countries. Technological factors such as the adoption of advanced manufacturing technologies (AMTs) can also affect unemployment rates. AMTs can lead to job displacement in traditional manufacturing industries, but they can also create new high-skill jobs in the technology sector.

While these factors can lead to increased productivity and efficiency, they can also lead to job displacement and unemployment, particularly in sectors such as manufacturing and agriculture. A study by Bristy (2023) found that automation and digitalization were significant predictors of unemployment rates in Bangladesh. One important economic factor that can impact unemployment rates is foreign direct investment (FDI). FDI can provide new job opportunities and boost economic growth, but it can also lead to job displacement and exacerbate inequality. A study by Ofori and Asongu (2021) found that FDI inflows were positively associated with unemployment rates in African countries, suggesting the need for policies that promote inclusive growth.

Another social factor that can affect unemployment rates is gender inequality. Women are often more vulnerable to unemployment and underemployment, and gender disparities in education and employment opportunities can limit their access to quality jobs. A study by Baffour and

Quartey (2016) found that gender inequality was a significant predictor of unemployment rates in Ghana, highlighting the importance of promoting gender equality in employment policies.

One potential policy intervention is investment in education and training programs to equip workers with the necessary skills to adapt to new technologies and job requirements. This can include vocational training programs, apprenticeships, and initiatives to promote lifelong learning. A study by Jatav and Sen (2013) found that education was a significant predictor of employment outcomes in India, highlighting the importance of investing in human capital development. Potential policy intervention is promoting entrepreneurship and small business development. This can create new job opportunities and diversify the economy, reducing reliance on traditional industries that may be susceptible to job displacement due to new technologies. A study by Li et al. (2021) found that entrepreneurship was positively associated with employment outcomes in Chinese cities, suggesting the potential for entrepreneurship policies to promote job creation.

Additionally, policies that promote inclusive growth and reduce inequality can also help mitigate the negative impact of job displacement and unemployment. This can include social safety nets, progressive taxation systems, and initiatives to promote gender equality and access to education and employment opportunities. A study by Chibba and Luiz (2011) found that redistributive policies were effective in reducing poverty and inequality in South Africa. However, it is important to note that policy interventions may have trade-offs and unintended consequences. For example, policies that promote entrepreneurship and small business development may also lead to greater competition and consolidation in certain industries, which can reduce job opportunities in the long run. Therefore, careful evaluation and monitoring of policy interventions is necessary to ensure they are effective and equitable.

Overall, understanding and addressing the unemployment rate in developing countries during the implementation of Industry 4.0 requires a multi-faceted approach that considers the complex interplay between economic, social, and technological factors. Policy interventions that promote human capital development, entrepreneurship, and inclusive growth can help mitigate the negative impact of job displacement and unemployment, but careful evaluation and monitoring are necessary to ensure these policies are effective and equitable. A comprehensive understanding of the key predictors and factors that contribute to unemployment rates in developing countries during the implementation of Industry 4.0 can help policymakers and stakeholders develop effective strategies for promoting job creation and reducing unemployment. Further research can also explore the potential trade-offs and synergies between different policy interventions and their impact on employment outcomes.

The key predictors or factors that contribute to unemployment rates in developing countries during the implementation of Industry 4.0 include economic, social, and technological factors. Understanding the interplay of these factors and their impact on employment can help policymakers and researchers develop effective strategies for promoting job creation and reducing unemployment in developing countries. Finally, it is important to note that the impact of these predictors and factors can vary depending on the specific context and characteristics of the developing country. For example, a study by Zhou and Tyers (2019) found that the impact of automation on unemployment rates in China was moderated by regional economic development and education levels.

**The ethical considerations that need to be taken into account when developing predictive models for unemployment rates**

The use of machine learning algorithms for predicting unemployment rates in developing countries during the implementation of Industry 4.0 raises important ethical considerations that need to be taken into account. These considerations include:

- **Bias and discrimination:** As mentioned earlier, machine learning algorithms are only as good as the data they are trained on. Machine learning algorithms can be prone to bias, which can have negative consequences for individuals and groups. It is important to be aware of the potential for bias and to take steps to mitigate it, such as using diverse and representative data sets, regularly testing and evaluating the models for bias, and making adjustments as necessary. If the data contains biases or discriminatory patterns, the algorithms can perpetuate and amplify these biases. This can lead to unfair and discriminatory outcomes, particularly for vulnerable or marginalized populations. Therefore, it is important to ensure that the data used to train these models is representative and free from biases.
- **Privacy:** The use of personal data in developing predictive models for unemployment rates can raise concerns about privacy. This is especially important in developing countries where data protection laws may be weak or non-existent. Therefore, it is important to ensure that appropriate measures are in place to protect the privacy of individuals whose data is being used.

- **Transparency and explainability:** Machine learning algorithms can be complex and difficult to understand. Therefore, it is important to ensure that these algorithms are transparent and explainable so that stakeholders can understand how the algorithms work and how decisions are made.
- **Accountability:** The use of predictive models for unemployment rates can have significant consequences for individuals and society as a whole. Therefore, it is important to ensure that there is accountability for the decisions made by these models. This can include mechanisms for redress and oversight.
- **Inclusivity:** The development of predictive models for unemployment rates should be inclusive and involve stakeholders from diverse backgrounds. This can help ensure that the models are representative and do not perpetuate existing power imbalances.
- **Informed consent:** The use of personal data in predictive models for unemployment rates raises concerns about informed consent. Individuals have the right to know how their data is being used and to have a say in how it is used. It is important to ensure that individuals are fully informed about the use of their data and that they have given their consent for it to be used.
- **Fairness:** Predictive models for unemployment rates should be designed to promote fairness and reduce inequalities. This means ensuring that opportunities are available to all individuals regardless of their race, gender, socioeconomic status, or other factors. It also means being aware of the potential impact of the models on different groups and taking steps to mitigate any negative effects.
- **Continuous evaluation:** Predictive models for unemployment rates should be continuously evaluated to ensure that they remain accurate, fair, and effective. This includes monitoring their impact on individuals and society as a whole and making adjustments as necessary.
- **Social responsibility:** The development of predictive models for unemployment rates should be guided by a sense of social responsibility. This means ensuring that the models are developed in a way that benefits society and promotes the public good. It also means being aware of the potential negative consequences of the models and taking steps to mitigate these risks.
- **Usefulness:** Predictive models for unemployment rates should be developed and used in a way that is useful and relevant to stakeholders. This means ensuring that the models are designed to address specific challenges and that they provide actionable insights that can be used to improve policy and decision-making.

By taking these ethical considerations into account, it is possible to use a predictive conceptual model for understanding and addressing unemployment rates contributing factors that are not only accurate and effective but also ethical, fair, and beneficial to society as a whole. In summary, the development of a predictive conceptual model for understanding and addressing the factors that contribute to unemployment rates in developing countries using machine learning algorithms during the implementation of Industry 4.0 raises important ethical considerations that need to be taken into account. These ethical considerations contribute to the achievement of sustainable and equitable development and the application of machine learning algorithms in understanding and addressing the unemployment rate at the time of Industry 4.0 implementation.

**Contributions**

The contributions to the body of knowledge for developing a predictive conceptual model for addressing the factors that contribute to unemployment rates in developing countries during the implementation of Industry 4.0 using machine learning can include:

- **Improved understanding of the relationship between Industry 4.0 implementation and unemployment rates in developing countries:** By using a predictive conceptual model, accurate forecasting, understanding, and addressing unemployment rates during Industry 4.0 implementation involves relevant variables. The proposed model provides a deeper understanding of the factors that affect employment in developing countries.
- **Identification of key drivers of unemployment in developing countries:** The use of a predictive conceptual model identifies the key drivers of unemployment in developing countries, such as education levels, economic growth, and technological advancements.
- **Creation of decision support tools for policymakers:** The predictive conceptual model provides decision support tools for policymakers, enabling them to make informed decisions on how to mitigate unemployment and promote economic growth when implementing Industry 4.0.
- **Improvement of existing models:** The predictive conceptual model contributes to the improvement of existing models, such as the Phillips Curve, which is widely used in macroeconomic analysis.
- **Advancement of machine learning techniques:** The predictive conceptual model facilitated by machine learning can contribute to the advancement of machine learning techniques and their application in economics and social sciences.

Overall, the contributions to the body of knowledge from a proposed predictive conceptual model in understanding and addressing the factors that contribute to unemployment rates in developing countries during the implementation of Industry 4.0 using machine learning can help improve academic understanding of the factors that affect employment and promote economic growth in developing countries.

**Steps in applying the predictive conceptual model**

Steps in applying the predictive conceptual model in understanding and addressing factors that contribute to unemployment rates using machine learning algorithms in developing countries. This study provides comprehensive steps of how a predictive conceptual model might be used through machine learning algorithms:

- **Identify the problem:** The first step in applying a predictive conceptual model is to clearly define the problem to be solved. In this case, the problem is predicting and addressing unemployment rates in developing countries during the implementation of Industry 4.0.
- **Collect data:** The next step is to gather data that can be used to train the machine learning algorithms. This might include data on economic indicators, such as GDP and inflation, as well as data on social and demographic factors, such as population growth and education levels.
- **Preprocess the data:** Once the data has been collected, it must be preprocessed to ensure that it is clean, accurate, and formatted correctly for use in the machine learning algorithm. This might involve removing missing values, normalizing the data, and splitting the data into training and testing sets.
- **Select a machine learning algorithm:** Many different machine learning algorithms can be used for predictive modeling, including Random Forest, Gradient Boosting, Support Vector Regression, and others. The choice of algorithm will depend on the specific problem being addressed and the characteristics of the data.
- **Train the model:** Using the training data set, the machine learning algorithm is trained to identify patterns and relationships in the data that can be used to predict unemployment rates in developing countries during the implementation of Industry 4.0.
- **Evaluate the model:** Once the algorithm has been trained, it must be evaluated to ensure that it is accurate and reliable. This might involve using a variety of performance metrics, such as mean squared error or R-squared, to assess the algorithm's performance.

- **Deploy the model:** Finally, the predictive model can be deployed in real-world applications to help policymakers and businesses make more informed decisions related to economic development, job creation, and poverty reduction in developing countries. In this study, Figure 1 conceptual framework was developed as a predictive conceptual model for understanding and addressing factors that contribute to the unemployment rate while implementing Industry 4.0 in developing countries through the application of a machine learning algorithm.

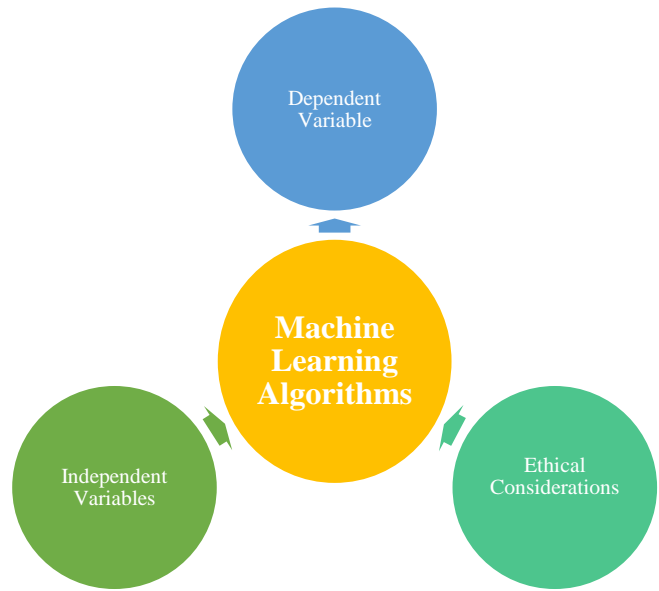

**Figure 1:** Predictive Conceptual Model

| Concept | Definition |
| --- | --- |
| Dependent variable | Unemployment rate |
| Independent variable | Economic Factors, Technological Factors, Political Factors, Social Factors |
| Machine Learning Algorithms | Regression, Decision Trees, Random Forest, Neural Networks |
| Ethical Considerations | Transparency, Privacy, Bias, Accountability, Usefulness |

**Table 1:** Predictive Conceptual Model

Table 1 presents a comprehensive summary of the variables that the constitute predictive conceptual model. This conceptual model identifies the key concepts and variables that need to be considered when using machine learning to predict unemployment rates in developing countries during the implementation of Industry 4.0. The model includes the dependent variable of the unemployment rate and several independent variables that may contribute to unemployment rates in the context of Industry 4.0 implementation. The economic, technological, political, and social factors are identified as potential independent variables. Machine learning algorithms, such as regression, decision trees, random forests, and neural networks, can be used to analyze the relationship between the dependent and independent variables. Finally, the framework includes ethical considerations that need to be taken into account when using machine learning algorithms to predict the unemployment rate. Here are a few possible ways that the predictive conceptual model (Figure 1) might benefit stakeholders:

- **Inform policy decisions:** Policymakers can use the model to inform decisions related to economic development and job creation in developing countries. By understanding how different economic and social factors are likely to affect unemployment rates, policymakers can make more informed decisions about where to invest resources and how to prioritize different policies.
- **Guide business decisions:** Businesses operating in developing countries can also use the model to guide their decisions related to hiring, expansion, and investment. By understanding how economic and social factors are likely to impact unemployment rates, businesses can make more informed decisions about where to locate their operations, what skills to prioritize when hiring, and how to respond to changes in the economic environment.
- **Forecast trends:** The model can also be used to forecast trends in unemployment rates in developing countries. By analyzing historical data and identifying patterns and relationships, the model can be used to predict future trends in unemployment rates, which can help businesses and policymakers prepare for future challenges and opportunities.
- **Monitor progress:** Finally, the model can be used to monitor progress in reducing unemployment rates in developing countries. By comparing predicted unemployment rates to actual rates over time, policymakers and businesses can track progress toward their goals and make adjustments to their strategies as needed.

Overall, the implementation of the predictive model depends on the specific context and the needs of the stakeholders involved. However, by leveraging the power of machine learning algorithms to predict unemployment rates in developing countries during the implementation of Industry 4.0, policymakers and businesses can make more informed decisions that can help promote economic growth, job creation, and poverty reduction.

**The implications, recommendations, and future study**

**Implications**

This model provides implications, first, it can help policymakers make informed decisions to mitigate unemployment and promote economic growth. Second, it can help identify the key drivers of unemployment in developing countries, such as education levels, economic growth, and technological advancements. Finally, it can advance understanding of machine learning algorithms and their application in economics and social issues that can help the implementation of Industry 4.0 in developing countries.

**Recommendations**

Based on the findings of the study, the following recommendations can be made:

- Policymakers should prioritize investment in education and training to equip the workforce with the skills required for Industry 4.0.
- Governments should encourage and support the adoption of Industry 4.0 technologies by providing incentives to companies and startups.
- Researchers should continue to develop and improve predictive models for unemployment rates in developing countries using machine learning, incorporating new data sources, and exploring new techniques.

**Future study**

There is still much to be explored in the area of developing predictive models for unemployment rates in developing countries during the implementation of Industry 4.0 using machine learning algorithms. Some potential avenues for future study include:

- Exploring the impact of different Industry 4.0 technologies on employment in developing countries, and how they can be harnessed to create jobs.
- Investigating the social and economic implications of Industry 4.0 on developing countries beyond just unemployment rates, such as income inequality and access to resources.
- Examining the role of government policies and institutions in promoting the adoption of Industry 4.0 technologies and mitigating their negative impacts on employment.
- Exploring the effectiveness of different machine learning algorithms in predicting unemployment rates in developing countries and identifying the most suitable ones for this task.

Overall, further research in this area can help improve academic and non-academic understanding of the complex relationship between Industry 4.0, employment, and economic development in developing countries, and inform policy decisions to promote inclusive growth.

**Conclusion**

The use of machine learning algorithms to predict unemployment rates in developing countries during the implementation of Industry 4.0 is a promising area of research with the potential to inform important decisions related to economic development, job creation, and poverty reduction. By analyzing economic and social factors, such as GDP, inflation, population growth, education levels, and technological advancements, researchers can develop predictive models that accurately forecast changes in unemployment rates in developing countries. Despite the challenges associated with data availability, model accuracy, and ethical considerations, the benefits of this research are significant. Policymakers and businesses can use the model to understand factors and variables to be considered when implementing Industry 4.0, it also provides the role of machine learning algorithms in understanding and addressing the unemployment rate when implementing Industry 4.0 in developing countries.

With continued research and development, policymakers and businesses can use this model to make more informed decisions that can help promote economic growth, job creation, and poverty reduction in developing countries.

**References**

1. Al Mamun, M. A., Ahmed, F., & Kamruzzaman, M. (2020). Artificial neural network modeling for forecasting unemployment rate: An empirical study in Malaysia. *International Journal of Emerging Markets*, 15(3), 515-531.
2. Albuquerque, P. C., Cajueiro, D. O., & Rossi, M. D. (2022). Machine learning models for forecasting power electricity consumption using a high dimensional dataset. *Expert Systems with Applications*, *187*, 115917.
3. Asian Development Bank. (2019). Industry 4.0 and the Future of Work in Developing Countries. Retrieved from https://www.adb.org/sites/default/files/publication/541236/industry-4-future-work-developing-countries.pdf
4. Baffour, P. T., & Quartey, P. (2016). A gendered perspective of underemployment in Ghana. *Ghana Social Science*, *13*(2), 209.

5. Bristy, N. I. (2023). *Mental Health Prediction Among Unemployed Graduates Using Machine Learning Approach: BD Perspective* (Doctoral dissertation, East West University).
6. Bughin, J., Hazan, E., Lund, S., Dahlström, P., Wiesinger, A., & Subramaniam, A. (2018). Skill shift: Automation and the future of the workforce. *McKinsey Global Institute*, *1*, 3-84.
7. Chang, F. J., Chang, L. C., & Liou, J. Y. (2021, December). Hybrid Machine Learning models with Principal Component Analysis for Multi-Step-Ahead Urban Flood Inundation. In *AGU Fall Meeting Abstracts* (Vol. 2021, pp. NH35F-09).
8. Chibba, M., & Luiz, J. M. (2011). Poverty, inequality and unemployment in South Africa: Context, issues and the way forward. *Economic Papers: A journal of applied economics and policy*, *30*(3), 307-315.
9. Chung, D., Yun, J., Lee, J., & Jeon, Y. (2023). Predictive model of employee attrition based on stacking ensemble learning. *Expert Systems with Applications*, *215*, 119364.
10. de Oliveira, L. G. M. (2023). Which One Predicts Better?: Comparing Different GDP Nowcasting Methods Using Brazilian Data.
11. Gabrikova, B., Svabova, L., & Kramarova, K. (2023). Machine learning ensemble modelling for predicting unemployment duration. *Applied Sciences*, *13*(18), 10146.
12. Gogas, P., Papadimitriou, T., & Sofianos, E. (2022). Forecasting unemployment in the euro area with machine learning. *Journal of Forecasting*, *41*(3), 551-566.
13. International Labour Organization. (2018). Future of Work in Industry 4.0. Retrieved from https://www.ilo.org/wcmsp5/groups/public/---dgreports/---cabinet/documents/publication/wcms_632489.pdf
14. Jatav, M., & Sen, S. (2013). Drivers of non-farm employment in rural India: Evidence from the 2009-10 NSSO Round. *Economic and Political Weekly*, 14-21.
15. Karasu, S., Altan, A., Bekiros, S., & Ahmad, W. (2020). A new forecasting model with wrapper-based feature selection approach using multi-objective optimization technique for chaotic crude oil time series. *Energy*, *212*, 118750.
16. Katris, C. (2020). Prediction of unemployment rates with time series and machine learning techniques. *Computational Economics*, *55*(2), 673-706.
17. Künzel, S. R., Sekhon, J. S., Bickel, P. J., & Yu, B. (2019). Metalearners for estimating heterogeneous treatment effects using machine learning. *Proceedings of the national academy of sciences*, *116*(10), 4156-4165.

18. Kurt, R. (2019). Industry 4.0 in terms of industrial relations and its impacts on labour life. *Procedia computer science*, *158*, 590-601.
19. Kütük, Y., & Güloğlu, B. (2019). Prediction of transition probabilities from unemployment to employment for Turkey via machine learning and econometrics: a comparative study. *Journal of Research in Economics*, *3*(1), 58-75.
20. Liu, L., Chen, C., & Wang, B. (2022). Predicting financial crises with machine learning methods. *Journal of Forecasting*, *41*(5), 871-910.
21. Mittal, M., Goyal, L. M., Sethi, J. K., & Hemanth, D. J. (2019). Monitoring the impact of economic crisis on crime in India using machine learning. *Computational Economics*, *53*, 1467-1485.
22. Mulaudzi, R., & Ajoodha, R. (2020, December). Application of deep learning to forecast the South African unemployment rate: a multivariate approach. In *2020 IEEE Asia-Pacific Conference on Computer Science and Data Engineering (CSDE)* (pp. 1-6). IEEE.
23. Nguyen, P. H., Tsai, J. F., Kayral, I. E., & Lin, M. H. (2021). Unemployment rates forecasting with grey-based models in the post-COVID-19 period: A case study from vietnam. *Sustainability*, *13*(14), 7879.
24. Ofori, I. K., & Asongu, S. (2021). Foreign direct investment, governance and inclusive growth in sub-Saharan Africa. *Governance and Inclusive Growth in Sub-Saharan Africa (June 7, 2021)*.
25. Park, S., & Yang, J. S. (2022). Interpretable deep learning LSTM model for intelligent economic decision-making. *Knowledge-Based Systems*, *248*, 108907.
26. Taira, K., Hosokawa, R., Itatani, T., & Fujita, S. (2021). Predicting the number of suicides in Japan using internet search queries: vector autoregression time series model. *JMIR public health and surveillance*, *7*(12), e34016.
27. Xu, W., Li, Z., Cheng, C., & Zheng, T. (2013). Data mining for unemployment rate prediction using search engine query data. *Service Oriented Computing and Applications*, *7*, 33-42.
28. Zainun, N. Y. B., Rahman, I. A., & Eftekhari, M. (2010). Forecasting low-cost housing demand in Johor Bahru, Malaysia using artificial neural networks (ANN). *Journal of Mathematics Research*, *2*(1), 14.
29. Zhou, Y., & Tyers, R. (2019). Automation and inequality in China. *China Economic Review*, *58*, 101202.

30. Zhu, H., Zhao, Y., Pan, Y., Xie, H., Wu, F., & Huan, R. (2021). Robust heartbeat classification for wearable Single-Lead ECG via extreme gradient boosting. *Sensors*, *21*(16), 5290.